# Digital Labor and the Inconspicuous Production of Artificial Intelligence

Antonio A. Casilli

## Abstract:

*Digital platforms capitalize on users' labor, often disguising essential contributions as casual activities or consumption, regardless of users' recognition of their efforts. Data annotation, content creation, and engagement with advertising are all aspects of this hidden productivity. Despite playing a crucial role in driving AI development, such tasks remain largely unrecognized and undercompensated. This chapter exposes the systemic devaluation of these activities in the digital economy, by drawing on historical theories about unrecognized labor, from housework to audience labor. This approach advocates for a broader understanding of digital labor by introducing the concept of "inconspicuous production." It moves beyond the traditional notion of "invisible work" to highlight the hidden elements inherent in all job types, especially in light of growing automation and platform-based employment.*

## INTRODUCTION

Modern[1] concerns over the vanishing nature of work reflect a deeper transformation of human occupations into contingent and invisibilized digital labor. Work carried out on apps, over the internet or within digital environments, often involves content generation, data annotation, and algorithm evaluation, among other activities. These occupations are characterized by taskification and datafication and encompass a broad spectrum of digital environments. On-demand platforms, sometimes referred to as gig-work platforms, connect freelancers directly with possible clients and with short-term jobs or gigs, primarily in transportation, delivery, and various freelance services, including those found on Fiverr and Upwork. Crowdsourcing and micro-task platforms aggregate small and more fragmented assignments, requiring seconds or minutes to complete, such as image tagging or micro-surveys, and allocate them to anonymous workforce, as seen on Figure 8 and Scale AI. Meanwhile, content creation and monetization platforms, such as YouTube and TikTok, enable a select group of creators to earn income through advertisements, subscriptions, or donations. Virtually everyone else contributes videos, texts, and images without compensation, and some are underpaid for performing basic tasks like data filtering, moderation, and engaging with online content. This variety of types and examples illustrates how platform labor is driven by the fragmentation of work into discrete tasks accessible to individuals globally.

Fragmentation into tasks and datafication are critical for contemporary artificial intelligence (AI). Machine learning models evolve by identifying patterns in data generated, sorted, and annotated by workers on digital labor platforms. On-demand platforms like Uber, microwork services like Amazon SageMaker GroundTruth, and social media like Instagram are pivotal in generating data that fuels AI solutions. For instance, Uber's integration of matching algorithms, GPS routing, and autonomous vehicles is achieved by collecting massive data of both passengers and drivers (Pidoux et al., 2024). Meta's use of public posts on Instagram and Facebook to enhance AI's understanding of human emotions and interactions (Loh, 2023), as well as Amazon's use of live user conversations to train Alexa (Zhou, 2023), are all examples of controversial methods of leveraging of users' digital labor to improve automated processes. The emergence of generative AI has intensified this process (Perrigo, 2023).

This chapter explores the role of digital labor performed by users of digital platforms, highlighting its significance within a data-driven economic landscape that often fails to adequately recognize or compensate this input. Blending labor theories from pre-digital and digital ages, it emphasizes the infrastructural elements of datafication and taskification in value creation. Platforms influence human economies by reshaping global supply chains and orchestrating extensive labor outsourcing, leveraging global wage disparities to tap into both local and international workforces, often at the expense of fair compensation and labor protections. Digital platform labor is often dismissed as negligible, blurring the lines between trivial tasks, leisure activities, freelancing, temping, piecework, and at times not acknowledged as "work" (see also Tran, Chapter 30). For several decades, social scientists have sought to shed light on labor that is too subtle to be recognized as a real job, even before the emergence of AI and digital platforms. They have highlighted how tasks usually considered too mundane and minor can have significant value-creating potential. By examining prominent theories of labor, I aim to argue for a broader understanding of "inconspicuous production," emphasizing its importance. Recognizing such labor is essential for reclaiming economic dignity, securing fair and improved working conditions, and ultimately facilitating social advancement for all workers and users of digital platforms.

---

[1] This text develops the arguments discussed in Chapter 3 "On-Demand Digital Labor" 6 "Work Outside Work" and 8 "Subjectivity at Work, Globalization, and Automation" of Casilli (2024).

## A GENEALOGY OF UNRECOGNIZED LABOR BEFORE THE PLATFORMS

Despite their diversity, digital platforms share a distinct trait: they simultaneously engage users in work while positioning them outside the sphere of work. As these platforms demand expertise and time from users, they relegate them to roles such as "participant," "consumer," "contributor," or "guest" – anything but "employee" or "staff." Thus, digital laborers struggle to gain recognition as workers who share in the benefits of standard employment. They are not alone. In this sense, they belong to a larger community of people who conduct unrecognized work. This includes domestic work (that relates to household and caregiving duties), gamified work (that features elements of play, such as competition or rewards), consumption work (that involves purchasing, using, or disposing of goods and services), audience labor (the act of producing value through being the public of media and target of advertisement), and immaterial labor (that produces goods and services by intellectual or creative activities). In recent decades, these categories have emerged, showing that platform workers' position, conditions, and activities are all affected by their degree of "conspicuousness" – meaning the extent to which their occupations are visible.

### Consumer Work

Historically overlapping with online user-generated content or early sharing economy where users' efforts drive platform success, contemporary digital labor, mirrors consumer work. This concept is not novel; tasks that were previously performed by employees, such as assembling furniture or printing tickets, have progressively been transferred to consumers, blurring the lines between customer and worker. This trend has escalated with the digital economy through innovations like self-service checkouts, emphasizing consumer involvement in labor. Today, in the context of digital platform labor, value creation relies increasingly on what some Ursula Huws (2003) calls the "unpaid labor of 'consumption work'" and Marie-Anne Dujarier (2015; 2016) dubs "consumer work."

Contemporary economies promise customization through consumer work: consumers personalize their own goods and services. Additionally, consumers communicate verbally or symbolically about the good they consume to improve its visibility or quality. However, consumer work is presented as something outside the market. This is what unites consumer work and digital labor. Platforms exploit "work outside of work," tapping into the "cheap excess capacities" at the fringes of the labor market, enabling customization and improving product visibility or quality through consumer engagement (Bauer & Gegenhuber, 2015). When dealing with users, digital platforms often avoid mentioning payment. Users are supposed to create data to be exchanged for content, computing power, entertainment, etc. The micropaid, underpaid, and unpaid activities that characterize digital labor are inextricably linked with "free consumer work."

Digital platforms, though, have shifted consumer work from an ancillary role to the core of a production system that involves a vast, enthusiastic, and compliant workforce at no cost. Social media platforms like Facebook and TikTok, for instance, harness users' voluntary content creation, engagement, and data generation as a form of unpaid digital labor. Similarly, in the video game industry, players contribute free labor through beta testing, modding, and audience interactions on streaming platforms like Twitch to generate revenue outside traditional employment costs. This merger of consumption and production marks a new phase in capitalist accumulation, commodifying services in food production, housing, and personal care, alongside the exchange of intangible goods and the formation of online communities.

Even micro-tasks and freelancing platforms like Microworkers and Fiverr offer standardized tasks for minimal compensation (and in some cases for no compensation at all, as the obstacles to remuneration on these services are often insurmountable for workers). But similar activities are also performed by consumers outside these platforms, contributing to the production of cultural content, social connections, and sometimes of annotated data for AI. For example, reCAPTCHA differentiates human users from bots through free tasks like reading text or identifying objects in images. Like in consumer work, everyday activities are harnessed as productive labor without explicit recognition or compensation. But in this case, this work also contributes to AI training, since these verification tasks produce data that enhance image recognition for Google Images, OCR algorithms for Google Books, or improve self-driving car technology for Waymo.

## Housework

Digital labor activities often incorporate significant elements of care work, personal services, and other roles typically linked to workforce reproduction rather than production. This type of labor, mirroring domestic and parental work, has historically been associated with societal expectations and stereotypes of women's roles (Dalla Costa & James, 1972; Jarrett, 2014; Schwartz Cowan, 1985), leading some to describe it as "digital housewife" (Jarrett, 2015) or "digital care work." Regardless of gender, every user assumes a modern version of a "housewife," engaging in essential, yet mostly unpaid, reproductive work online. The nature of digital labor blurs the lines between private and public spheres, reproduction and production, and the concepts of use value and exchange value. Research highlights the tendency to overlook both domestic and digital work, pointing to a systemic undervaluation and invisibility, thus perpetuating a form of domestic servitude by disguising user involvement on digital platforms as social participation (Menking & Erickson, 2015).

The integration of generative AI, such as OpenAI's GPT models, into this framework does not escape the dynamics of exploitation, particularly during its training and reinforcement phases. These phases are marked by "affective turbulence" among users, especially when biases in AI output affect marginalized communities, exacerbating intersectional concerns (Perrotta et al., 2022).

Digital labor, while engaging users in creative and communal activities, perpetuates existing societal values and hierarchies. Feminist critiques have long challenged the ideological mystification surrounding unpaid labor, particularly housework (Dalla Costa, 2008), arguing that it imposes an exploitative labor relation under the guise of "love" (Jarrett, 2003) or "friendship" in digital settings. This discourse extends to demanding remuneration for the digital contributions of internet users (like in Laurel Ptak's (2014) Wages for Facebook), drawing parallels to historical movements advocating for the recognition and compensation of housework (Federici, 1975).

## Audience Labor

A different approach to unrecognized labor was initiated by Dallas Walker Smythe (1977; 1978). Smythe's concept of "audience commodity" challenged the traditional view of media consumption by considering the time spent by audiences on commercial media as a form of labor that contributes significantly to value creation. This labor, according to Smythe, turns audience members into producers of "ideology and consciousness," serving advertisers and marketers. While developed for traditional media, Smythe's analysis applies to digital platforms. Free newspapers and commercial radios from the 20th century, supported by advertisements and available at no cost to readers and listeners, served as precursors to today's search engines and streaming platforms: advertisers are charged, users access free content, and creators receive compensation. Smythe interprets this free access as the audience creating value for media owners and advertisers without direct compensation, challenging the notion of "free" consumption, which operate as two-sided markets where advertisers pay, but the audience often contributes value without direct compensation.

The transition from audience labor to digital labor highlights two key distinctions: the type of activities and their purpose. Unlike traditional audience labor focused solely on consumption, digital labor includes creating online presence and content, essentially performing the "work of being watched" and managing social indicators (Andrejevic, 2002). This dual role underscores the exploitation of digital labor, as it not only qualifies objects for monetization but also involves tasks like data labeling and image recognition, crucial for training algorithms and enhancing machine learning processes. Recent projects like Google DeepMind's use of StarCraft II gameplay for training its AlphaStar AI model (Jaderberg et al., 2019) and OpenAI's application of reinforcement learning across various services, from ChatGPT to Dall-E illustrate how user interactions contribute to significant technological advancements.

### Immaterial Labor

The concept of immaterial labor, as introduced by philosopher Maurizio Lazzarato in the 1990s, designates the work carried out in cultural industries aimed at adding value, sharing, and recommending content within the scope of cognitive capitalism (Lazzarato, 2006). Like other notions discussed supra, immaterial labor blurs the lines between productive and reproductive work, encompassing cultural, relational, and cognitive activities as well as time itself as market commodities. It focuses on the creation of value not through the tangible transformation of reality but by enhancing a commodity's informational content, often through unrecognized activities like participation on social media or microwork aligning data with human cultures and values.

Immaterial labor is distinct from digital labor, primarily because it is considered an "intellectual" labor that aligns with the Marxian concept of the "general intellect," which involves occupations requiring a high level of creativity. This type of labor is performed by what Paolo Virno (2001) terms "mass intellectuality," a group characterized by cognitive competences that resist objectification in machinery. Unlike the repetitive tasks associated with digital labor and automation, immaterial labor involves creative and intellectual efforts. The most visible traditional examples of immaterial workers included those in marketing, audiovisual production, advertising, fashion, and photography, contrasting with today's digital sector roles like software and multimedia content production. These fields, considered part of the advanced tertiary sector, often involved specialized professions with standard employment contracts, though subcontracting and self-employment were also common. This historical perspective highlights an "exalted" vision of immaterial labor, often overshadowing the less visible, yet significant, contributions of digital platform users who share several traits with traditional "material" labor. The stark distinctions between physical and intellectual, manual and skilled labor that were already challenged in the 1970s (Sohn-Rethel, 1978) have been further undermined in the digital age, as demonstrated by studies in internet culture and the philosophy of science (Pasquinelli, 2023; White, 2006). The daily activities of a lawyer or warehouse worker are as concrete as those undertaken by users of online platforms. This includes click farmers in Venezuela, Facebook moderators in Kenya, data annotators in India, and Uber drivers in Germany, all of whom face tangible challenges and engage physically with their work, utilizing their bodies, locations, and senses. This perspective aligns with other scholarly work on digital labor (Dyer-Witheford, 2015; Bulut, 2020), challenging the notion of its purely immaterial nature. Instead, it posits that the tangible and cognitive elements of platform-based work are inextricably linked and equally significant, offering a more nuanced understanding of this contemporary form of labor.

## DISCOVERING INCONSPICUOUS DIGITAL LABOR

Examining the earlier concepts of consumption work, housework, audience labor, and immaterial labor, I have attempted to pinpoint the features of contemporary digital labor, while resolving some of the dichotomies (between work and non-work, tangible and intangible) that hinder a full comprehension of the contribution of users to digital platforms. The analysis of these notions reveals, however, that there is a more fundamental contrast between forms of work that are readily identifiable as such and activities that are mediated by digital technologies and transformed into "silent," inconspicuous, or simply "implicit" tasks through the use of AI and algorithmic tools.

## Work Invisibility as Mirror of Job Instability

It is difficult to make work both visible and valuable, as exemplified by the lack of visibility of domestic labor, by the casual nature of consumer and audience labor, and the immaterial aspects of cognitive capitalism. The problem of recognition applies to all types of occupations. Actually, according to philosopher Axel Honneth, recognition occupies a central place in social and subjective life (Honneth, 2015 [1992]). The development of AI solutions, recommendation algorithms, and search engines depends on the work of users on digital platforms who seek acknowledgment on both social and political levels.

According to sociologists Susan Leigh Star and Anselm Strauss's (1999) seminal contribution, "Layers of Silence, Arenas of Voice" worker visibility can be enhanced by highlighting and describing their methods of operation, their professional accomplishments, and the challenges that require their specialized knowledge. In contrast, when these elements are silenced, the occupation is often not recognized as "real work." It is in this oscillation between these two poles that a definition of what exactly constitutes work at a given time and in a given society emerges. Many activities that are not self-evidently work may be ambiguous in the sense that they cannot be labeled as such, as we have seen in the case of housework and care. A given society's recognition of what constitutes work can be traced back to long historical processes of professionalization, which, made visible, can help create the status of a career in a competitive market. Conversely, it is possible for the definition of what constitutes work to change in such a way that certain activities previously classified as work are gradually deprived of their status and silenced.

Considering the many transformations that have taken place in the global labor markets since Star and Strauss were writing in the last quarter of the 20th century, this way of envisioning work is all the more relevant. Recent scholarship has challenged the traditional concept of invisible work, arguing it stems from a Western-centric view that oversimplifies and naturalizes historically complex power dynamics (Raval, 2021). Interdisciplinary research in digital labor, HCI, and STS has revealed that invisibility in affluent nations is a nuanced construct. Moreover, these studies question the assumed "public" to whom work is invisible, suggesting a more complex dynamic than initially proposed by Star and Strauss. Moreover, invisibility in wealthier countries proves to be a complex construction rather than an intrinsic quality of work. This is due to both technological and economic factors.

Even in high-income countries, structural unemployment, waves of business restructuring, and concerted outsourcing policies have diminished the prospect of employment stability. For new employees both in the North and in the Global South, the prospect of obtaining a secure position has diminished. Subcontracting trends and the increase in nonstandard forms of employment have created a climate of general uncertainty regarding where and what work is. A rich body of literature dating from the late 1990s to the present has described the volatility, fragmentation, and riskiness of labor during the same period (Bauman, 1998; Beck, 2000; Sennett, 1998; Weil, 2017). The shift to remote work, especially highlighted during the global COVID-19 pandemic, coupled with a series of layoffs, some of which are blamed on GenAI but likely stem from platforms' pressures to turn a profit, has accentuated certain trends. These developments are part of a broader movement toward the proliferation of nontraditional employment regimes. According to the International Labor Organization, they include temporary jobs, part-time work, multiemployer work, and concealed work. On-demand gigs, task-based work, and variable earnings from content creation all exhibit characteristics that align with this definition.

Companies are leveraging these atypical employment models to streamline their workforce, reduce payroll expenses, or engage those who prefer not to commit to full-time or permanent roles. The proportion of individuals employed under these arrangements fluctuates considerably, ranging from 23 to 40%, depending on whether the context is Europe, the US, or lowerand middle-income countries. In 2021, 89% of companies listed in the S&P 500 (indicative of the largest market capitalization firms on American exchanges) disclosed some form of contractor use in their yearly financial statements (US Government Accountability Office, 2022). Globally, there's an ongoing shift from uniform staff structures to a "blended workforce," where permanent employees and on-demand workers from digital platforms collaborate more closely (Mahato et al., 2021). This movement toward more flexible, casual, and temporary work has blurred the boundaries between phases of a career or moments in a day where one performs activities that "count as work" and phases where work becomes invisible and resembles casual communication, information consumption, or entertainment.

## From Invisible Labor to Inconspicuous Production

Platform-based digital labor accelerates these developments, making human productive activities fragmented and difficult to categorize, leading some observers to prematurely claim "the end of work" (Rifkin, 1995). The challenge is analyzing diverse, unstable, nonstandard jobs (like, for instance, a celebrity influencer monetizing short videos or a worker of the global peripheries clicking on a smartphone to earn a few cents) and connecting them to broader economic sectors. Ignoring these trends not only masks how work is changing but also hides the tactics corporations use to reduce labor costs while promising ambitious automation programs to their investors.

Changing labor trends offer an opportunity to discuss what work entails. In light of the tensions in the digital labor debate and the ones emerging from the labor market, it is all the more essential to negotiate and decide which activities are acceptable as work within the ever-expanding gray area of nonstandard regimes of employment.

Invisibility is a notion that has been used to navigate this gray area. It is a significant theme in labor literature, a trend that has persisted since research in the 1980s challenged the "folk concept" of work as activities "you have to do" to earn payment, thereby characterizing invisible labor as private, unpaid, and intimate occupations (Kaplan Daniels, 1987). Subsequent scholarship, on the other hand, has advocated for a less essentialist view of invisible labor. This includes recognizing that work and workers exist in places where they were initially unseen, while also acknowledging that invisibility persists in contemporary economic settings. Traditional companies as well as new organizational forms are affected by this issue. A broad range of activities are included in invisible labor in venues such as large retail shops and shopping malls, restaurants, upscale retail stores, and nonprofit organizations (Ballakrishnen et al., 2019). Work exists "along a spectrum of visibilities" yet the position of individuals on this spectrum is frequently determined by economic, social, and cultural capital, and often negatively affected by exclusion on the basis of gendered, racialized, or disability-related factors (Crain et al., 2016).

If the definition of what counts as labor in a particular context can be shaped by deliberate strategies of exclusion, invisibility should be seen not as a mere occurrence, but as the result of specific actions and methods implemented by social actors. In the previously mentioned article, Star and Strauss note that a common strategy for erasing work, thereby rendering it silent and invisible, is to introduce “large-scale networked systems.” By changing how tasks are allocated, they create an illusion that machines are performing the work, when in reality, it is delegated to other, concealed, human beings. Transposing this analysis into the contemporary technological environment broadens the authors' insights. Whenever technology is layered onto work, some visible effects are credited to machines, leading to the devaluation and erasure of human labor. The algorithm of a ride-hailing app that assigns tasks to drivers also makes the coordination work and price negotiation process invisible. On e-commerce platforms, a recommendation system that proposes purchases to consumers, likewise hides the human work involved in checking prices, removing duplicates, and updating product descriptions. Similarly, large language models emphasize the quality of algorithms and the quantity of parameters, yet neglect the foundational efforts of microtaskers and users' contributions to pre-training and reinforcement learning.

On digital platforms, the distribution of what counts as work is therefore inseparable from the distribution of its visibility. Both distributions influence the definition and value of the data and content that circulate within informational infrastructures. A lack of recognition of work, for instance, can lead to a lack of visibility of data – and vice versa. For example, users are unaware of the efforts made by individuals who moderate content on social media, as they do not see the tags and flags that moderators apply to the content they review (Roberts, 2019). When examining the lack of recognition of digital labor through the perspective of the ratio of visible to invisible work, we gain deeper insight into the broader social context where individuals engaged in concealed tasks struggle to access the job market. The integration of automation, networking, and the alignment of business processes with information and communication technologies play significant roles in silencing and eradicating work.

However, analyses focusing on invisibility often overlook that platform labor is segmented into various tasks, with some systematically less visible than others. Uber Eats couriers manage visible deliveries alongside less noticeable tasks such as bike maintenance and communicating with customers via the app. On Amazon Mechanical Turk, crowdworkers carry out visible microtasks but also engage in background activities like training, researching, and managing payment complexities. YouTube moderators not only oversee visible content but also deal with emotional challenges behind the scenes. This limitation is partially mitigated by the link the authors make between “invisible work” and “shadow work.” This notion, introduced by social critic Ivan Illich, alludes to the “shady side of industrial economy” most apparent in “the preparation for work to which one is compelled” (1980, p. 8). Shadow work is not merely the shadow of what is considered “real” work, nor is it work that has been pushed into the shadows by conflicts over the definition of work itself. Instead, it indicates that a portion of each activity is hidden from sight.

Another perspective on digital platform labor emerges from the concept of conspicuousness. Within this notion, occupations can be understood as an interplay between elements that are alternately illuminated and shadowed, visible and invisible, resulting in a more dynamic perspective on labor. “Conspicuous consumption” was, of course, coined by the American sociologist Thorstein Veblen to express the social significance of the act of consumption, not to satisfy needs, but as a display for others and to signal one's social status. It was legal scholar Jethro Lieberman who derived direct inspiration from this concept to introduce its counterpart, “inconspicuous production” which describes the production process that is not a spectacle, but an unrecognized chore, that is nevertheless crucial to performing the job at hand (Lieberman, 1970). In numerous occupations, “unskilled chores” make up the majority of the daily work, regardless of whether it is carried out by a high-earning professional or a low-wage worker. The difference is sharp between the portion of a lawyer's occupation that consists in putting evidence in a file without thought, and the part that entails appearing before a judge in court. Likewise, a short order cook meticulously prepping ingredients stands in stark contrast to the same individual announcing the completion of a dish with a bell ring. This was true before the rise of digital platforms, as Lieberman explored a range of occupations such as barbers, morticians, dentists, and keepers of mental wards. He highlighted their search for a balance between the conspicuous and ritualized part of their work (meeting with clients and carrying out work in public), and another, unobtrusive and ordinary part, which was performed out of sight. In this pre-internet context, inconspicuous labor consisted mainly of routine, and sometimes arduous, tasks that could be complex and difficult to communicate.

Therefore, in principle, each occupation can be characterized by a ratio of conspicuous to inconspicuous work. This does not necessarily mean that the ratio is higher for creative and specialized professions or lower for clerical work, administrative tasks, or even gigs and microtasks. Lieberman, in keeping with the culture of his time, was convinced that routine tasks that represent the inconspicuous portion of any job "can be accomplished by anyone with a modicum of training if routine enough, in fact, it can even be adapted to the machine" (1970, p. 90). Still, those who design automated processes, algorithms, and robots will find it difficult to make machines do their tasks, particularly the elementary and commonsensical ones that AI cannot simulate. The more inconspicuous the production activity and the less formalized the know-how it requires, the more challenging it is to automate. Processes that deal with formal and symbolic reasoning or factual knowledge lend themselves more easily to automation. However, processes necessitating comprehension and potentially experience of both the context and the meaning do not. A machine learning model will more readily arrive at the answer to a formal question like "What is the natural logarithm of p" than to a common-sense one like "how many feet can fit in a shoe?" (Floridi & Chiriatti, 2020).

Our discussion does not imply that the nature of digital labor is best captured by obscure concepts from the late 20th century. Rather, one could argue that the manner in which authors described automation within the socio-technological contexts before the advent of digital platforms is only somewhat applicable today. Lieberman thought that machines would replace inconspicuous work. Star and Strauss blame large-scale technological systems with "silencing work and continuing invisibility."

Importantly, the concept of "inconspicuous production" aligns with core sociological insights into technology's role in fostering fragmentation and rationalization, resonating with the notions of "taskification" and "datafication" in digital labor studies. Harry Braverman's seminal work Labor and Monopoly Capital (1974) argues that capitalism engenders a systematic degradation of work through deskilling and fragmentation. He posits that an expanding managerial class, responsible for work process design, creates a chasm between themselves and the workers relegated to mere execution, resulting in a loss of worker autonomy and task complexity. Predating Braverman, Georges Friedmann's 1956 classic The Anatomy of Work offers a complementary critique of labor fragmentation in industrial societies. Friedmann contends that increasing division of labor and mechanization lead to a "crumbling" of work, aptly captured in the French title's reference to "work in crumbs" (Le travail en miettes). This splintering, he argues, divorces workers from the final product, diminishing their sense of meaning and satisfaction.

The concept of inconspicuous production in digital labor echoes earlier theories of work degradation and splintering. Modern digital platforms, while revolutionizing work processes, often exacerbate task concealment, further alienating workers from their labor. This parallel extends classic labor critiques into the digital era, illuminating how contemporary work structures persistently undermine worker autonomy and connection to end products or services. Moreover, the notion of inconspicuous production expands upon Gray and Suri's (2019) concept of "ghost work," which they specifically apply to uncredited data trainers in AI development.

Expanding on Illich and Lieberman's ideas, inconspicuous digital labor encompasses the pervasive taskification and datafication embedded in all human activities. In this regard, it differs from invisible work, which refers to entire undervalued occupations. Inconspicuous production describes subtle and unnoticed aspects present in all jobs. This distinction underscores the ubiquity of inconspicuous elements across the labor spectrum and suggests that the distribution between conspicuous and inconspicuous labor transcends particular types of work. It thus departs from previous analyses focused on gendered, media, or cognitive labor. Importantly, it unveils how platforms leverage hidden production to create an illusion of automation, offering a more nuanced understanding of labor dynamics in the digital era.

## CONCLUSION: LABOR AND THE SISYPHEAN REALITY OF AI

Discussions about the nature of inconspicuous production are not merely academic exercises but have direct consequences on our understanding and anticipation of future labor market dynamics. With inconspicuousness as a focus, a particularly thorny topic, job displacement as a result of technological innovation, can be addressed in a more nuanced way. While the concern that machines will replace human labor dates back to the beginning of the industrial revolution, recent studies on the impact of AI have highlighted one of the main fault lines in such studies: automation is restricted to only the ostensible components of each occupation.

Specifically, let us consider two landmark studies that were published a decade apart: Frey and Osborne's (2013) "The Future of Employment" and Eloundou et al.'s (2024) "GPTs are GPTs" authored by OpenAI and the University of Pennsylvania researchers. The former famously predicted that 47% of the US employment base would be lost to automation by 2030, a groundbreaking forecast that sparked widespread discussion regarding the future of work. OpenAI's study on the impact of generative AI on labor emphasizes the urgency of the situation even more. In the study, it is estimated that 80% of the US workforce will be affected by Large Language Models (LLMs) at least to a certain extent, while approximately 19% will be affected to a greater degree. Not only do these alarming estimates clearly aim at demonstrating the transformative power of AI and automation, but they also suggest a wave of unprecedented job displacement is expected, and that corporations and financial institutions should not be held accountable – that is simply the way technological progress functions.

There is a common approach underlying both studies, which relies on a task-based analysis. Jobs are divided into discrete tasks, and if the percentage of these tasks that can be codified into machine learning models exceeds a somewhat arbitrary performance threshold, then the entire occupation can be considered replaced by AI.

This approach inherently focuses on visible labor – those aspects of a job that are easily identifiable and quantifiable. Consequently, it equates visible labor with labor that is at risk of replacement by technology. For a task to be codified, it must be recognizable. Therefore, it is not surprising that the authors focus on the conspicuous aspects of each job. Additionally, they introduce a cynical syllogism in support of their argument: work consists only of conspicuous elements that can be automated; inconspicuous elements cannot be automated, thus they are not work. By concentrating exclusively on the tangible aspects of jobs, this methodology potentially underestimates the complexity of job transformation and overestimates the ease with which entire occupations can be automated. It underscores a significant limitation in current discourse on labor and technology, suggesting a need for more nuanced analytical frameworks that encompass both the visible and invisible dimensions of work.

These studies (as well as layoffs and automation plans predicated on the same premises) fail to take into account that, whenever algorithmic solutions, AI systems, and unmanned devices have been introduced in the last few years, this has systematically resulted in an exposition of the amount of work involved, despite the inconspicuousness of this work has made it undetectable. As opposed to the claim that AI is the ultimate labor-saving technology, recent scholarship in the field of digital labor has developed new concepts such as fauxtomation (Taylor, 2018) and heteromation (Ekbia & Nardi, 2017) that explain how automation does not eliminate or take away work; rather, it changes the persons performing it.

Furthermore, I would argue that it alters the proportion of conspicuous and inconspicuous tasks within each job, so that the former can be targeted by automation and the latter are made even less noticeable and, as much as possible, go unrecognized and uncompensated.

This attitude is not new, and actually is shared by the very pioneers of machine intelligence. Nils Nilsson (2005) proposed substituting the Turing test with the "employment test." The Turing test assesses a computer's sophistication by its capability to mimic human-like interactions to the extent that users cannot discern whether they are interacting with a human or a machine, marking the machine's achievement of "human-level intelligence." Rather than simulating human conversation, like in the Turing test or the long tradition of chatbots from ELIZA to ChatGPT, Nilsson's test focuses on replacing conspicuous, paid jobs, and on ignoring the unpaid and inconspicuous contributions of digital labor. His approach suggests that the advancement toward human-equivalent AI should be gauged by how many of these tasks machines can competently handle. The main measure of success in AI, according to this perspective, hinges on its ability to substitute for human roles in the workforce, especially those that are paid and recognized. Unseen digital tasks, for which humans tend not to get compensated, are not even considered: they lie outside the scope of the test.

A significant issue in the era of automation and platformization is the increasing proportion of tasks that fall under the category of inconspicuous work, critical to data production and machine learning supervision, but not sufficiently rewarded. The digital labor market sells outcomes to companies as automated, yet fails to acknowledge the workers behind these tasks provide essential contributions. This misrepresentation suggests that digital labor is only a temporary need, with the ultimate goal of machines learning to operate independently of human input.

However, the reliance on data and the need for constant updates to reflect contemporary social norms underscore a fundamental challenge: achieving complete automation may be an unattainable goal. As algorithms are trained on historical data, their ability to adapt to new situations without human intervention is limited. This ongoing need for human input not only perpetuates the involvement of digital laborers but also maintains a distance between AI capabilities and human intelligence.

The Sisyphean reality of present-day machine learning is that the more AI aims to automate human tasks, the more it depends on humangenerated data. This dependency illustrates that the path to a fully autonomous machine, one that could function without human-crafted input, remains elusive. If computing's destiny is to achieve perfected AI, it is inevitably entwined with digital labor.